\documentclass[prd,preprint,eqsecnum,nofootinbib,amsmath,amssymb,
               tightenlines,dvips]{revtex4}
\usepackage{graphicx}
\usepackage{bm}
\input epsf

\begin {document}


\def\Mrowczynski{Mr\'owczy\'nski}
\def\Mrow{\Mrowczynski}

\def\lsim{\mbox{~{\raisebox{0.4ex}{$<$}}\hspace{-1.1em}
	{\raisebox{-0.6ex}{$\sim$}}~}}
\def\centerbox#1#2{\centerline{\epsfxsize=#1\textwidth \epsfbox{#2}}}
\def\ca{C_{\rm A}}
\def\cf{C_{\rm F}}
\def\da{d_{\rm A}}
\def\df{d_{\rm F}}

\def\tf{t_{\rm F}}
\def\nf{N_{\rm f}}
\def\Eq#1{Eq.~(\ref{#1})}
\def\alphas{\alpha_{\rm s}}

\def\k{{\bm k}}

\def\third{{\textstyle{\frac13}}}

\def\sixth{{\textstyle{\frac16}}}

\def\feq{f_{\rm eq}}
\def\Peq{P_{\rm eq}}
\def\S{{\cal S}}

\def\u{u}
\def\uvec{{\bm u}}
\def\mo{m_0}
\def\Mo{M_0}
\def\moa{m_{0,a}}

\def\p{{\bm p}}

\def\x{{\bm x}}

\def\v{{\bm v}}

\def\grad{{\bm\nabla}}
\def\da{d_{\rm A}}

\def\md{m_{\rm D}}



\title
    {
     The Bulk Viscosity of High-Temperature QCD
    }

\author{Peter Arnold and \c{C}a\=glar Do\=gan}
\affiliation
    {%
    Department of Physics,
    University of Virginia, Box 400714,
    Charlottesville, Virginia 22901, USA
    }%
\author{Guy D.\ Moore}
\affiliation
    {%
    Department of Physics,
    McGill University, 3600 University St.,
    Montr\'eal QC H3A 2T8, Canada
    }%

\date {\today}

\begin {abstract}%
    {%
      We compute the bulk viscosity $\zeta$ of high-temperature QCD to
      leading order in powers of the running coupling $\alphas(T)$.  We
      find that it is negligible compared to shear viscosity $\eta$ for
      any $\alphas$ that might reasonably be considered small.  The
      physics of bulk viscosity in QCD is a bit different than in scalar
      $\phi^4$ theory.  In particular, unlike in scalar theory, we find
      that an old, crude estimate of
      $\zeta \simeq 15 \big(\third-v_{\rm s}^2\big)^2 \eta$
      gives the correct order of magnitude, where
      $v_{\rm s}$ is the speed of sound.
      We also find that leading-log expansions of our result for $\zeta$
      are not accurate except at very small coupling.
}%
\end {abstract}

\maketitle
\thispagestyle {empty}


\section {Introduction and Results}
\label{sec:intro}

Studies of collective flow at RHIC \cite{collective_flow}, particularly
of elliptic flow, seem to be well described by nearly ideal
hydrodynamics \cite{hydro_people}.  In fact, it has recently been
claimed that these experiments prove that the quark-gluon plasma is the
most nearly ideal fluid known, with a viscosity close to the conjectured
lower bound on viscosities in any system \cite{claim_people,AdS_people}.
Such startling claims should be tested in any way we have available.
This requires studying flow in heavy ion collisions using non-ideal
hydrodynamics, that is, hydrodynamics including viscous effects
\cite{non_ideal_hydro}.  It also would be valuable to know
as much as possible about the theoretical expectations for viscosity in
the quark-gluon plasma.

In an ideal hydrodynamical treatment,
the evolution of the plasma is determined by
stress-energy conservation, $\partial_\mu T^{\mu \nu}=0$, together with
an equilibrium equation of state which
relates the pressure to the energy density,
$P = \Peq(\epsilon)$.
This should work whenever the system is locally in
equilibrium, which is the case in the limit of arbitrarily slowly varying flow
velocity $\u_i(x)$.  When $\u_i(x)$ varies somewhat in space,
the fluid will not be
precisely in local equilibrium, which will modify the stress tensor.
For slowly
varying $\u_i(x)$, the corrections to the stress tensor $T_{ij}$ can be
expanded in gradients of $\u_i$.  The leading order corrections are
parametrized by two quantities, the shear viscosity $\eta$ and the bulk
viscosity $\zeta$:
\begin{equation}
\label {eq:Tij}
T_{ij} =
\Peq(\epsilon) \, \delta_{ij} - \eta \left( \partial_i \u_j + \partial_j \u_i
- \frac{2}{3} \delta_{ij} \partial_k \u_k \right)
- \zeta \delta_{ij} \grad\cdot\uvec
\, ,
\end{equation}
where the expression is implicitly written in the instantaneous local
rest frame (where $T_{0i} = 0$).

While we are really interested in the viscosities $\eta$ and $\zeta$ of the
quark-gluon
plasma at temperatures $T \sim 200$ MeV, where the theory is far from
weakly coupled, we only possess reliable tools for computing dynamical
properties such as viscosities at weak coupling.%
\footnote%
    {%
    The lattice is a rigorous nonperturbative tool for studying
    thermodynamic properties of the quark-gluon plasma at strong
    coupling, but dynamical properties such as viscosities are hard to
    study on the lattice; see for instance,
    Refs.\ \cite{Aarts,Petreczky,Teaney}.
    }
Hopefully, extrapolating these results to strong coupling should give
the right ballpark for the same quantities at moderately strong coupling, with
uncertainties of perhaps a factor of a few.  This motivates
investigating $\eta$ and $\zeta$ at weak coupling.

In a relativistic system, on dimensional grounds, both $\eta$ and
$\zeta$ must scale as $\eta,\zeta \propto T^3$.  
A great deal of study has gone into
the shear viscosity in QCD.  It has been known for 20 years that
the parametric behavior is $\eta \sim T^3 / (\alphas^2 \log[1/\alphas])$
\cite{Hosoya,HosoyaKajantie};
the leading coefficient was closely estimated in 1990
\cite{Baym}, and complete results now exist
both at leading logarithmic order
\cite{lead-log} and full leading order \cite{lead-order} in the QCD
coupling $\alphas$.  On the other hand, the calculation of the bulk
viscosity has been completely neglected.
To our knowledge, no paper in the literature even correctly states what
{\em power} of $\alphas$ it is proportional to.  The purpose of this
paper is to fill this gap, by computing the bulk viscosity in weakly
coupled QCD at leading order in $\alphas$, using kinetic theory.
We will only consider the case of vanishing (or negligible)
chemical potential, $\mu = 0$.

In the next section, we will review the relevant physics of bulk
viscosity, explaining why the parametric behavior is
\begin{equation}
\zeta \sim \frac{\alphas^2 T^3}{\log[1/\alphas]}\quad (\mo\ll\alphas T)\,;
\qquad \qquad
\zeta \sim \frac{\mo^4}{T \alphas^2 \log[1/\alphas]} \quad
(\alphas T \ll \mo \ll T) \, .
\label{eq:param_result}
\end{equation}
Here $\mo$ refers to the heaviest zero-temperature
(current) quark mass which is smaller
than or of order the temperature $T$.
We use the subscript zero to emphasize that $\mo$ represents
a zero-temperature mass and not a finite-temperature effective
quasi-particle mass.
We will see that the physics of bulk viscosity is much
richer than that of shear viscosity.  In particular, the conformal anomaly
({\it i.e.}\/ scaling violations) and
the corrections to quasi-particle
dispersion relations due to interactions, both
irrelevant for shear viscosity, are both
essential pieces of physics for bulk viscosity.  Particle number
changing interactions also play a much larger role in bulk than in shear
viscosity.
These qualitative points have been anticipated by the pioneering
work of Jeon and Yaffe \cite{JeonYaffe,Jeon} on bulk viscosity in
relativistic $\phi^4$ theory.
However, we shall see later that there are
some significant qualitative
differences between bulk viscosity in $\phi^4$ theory
and in QCD.

\begin{figure}
\centerbox{0.7}{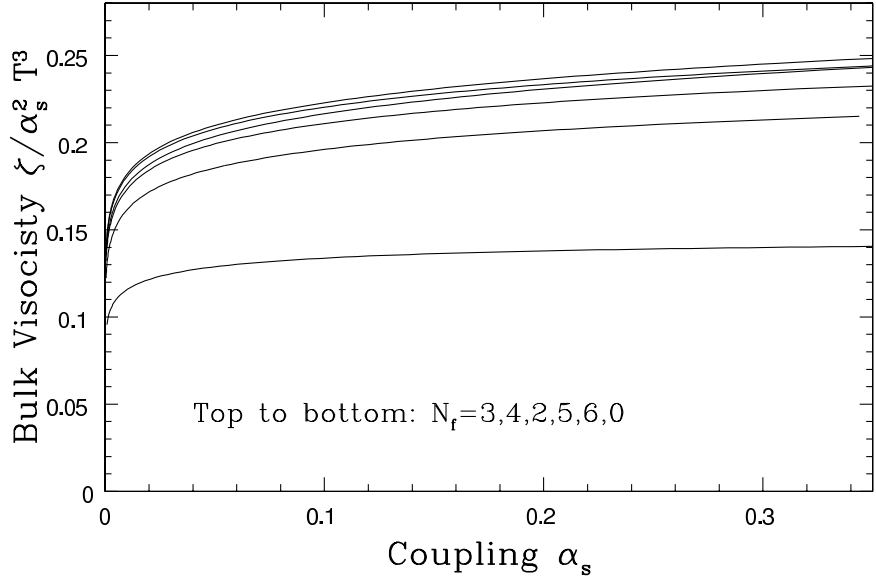}
\caption{\label{fig1}
Bulk viscosity for massless QCD at several values of $\nf$,
as a function of the coupling $\alphas$.
}
\end{figure}

\begin{figure}
\centerbox{0.7}{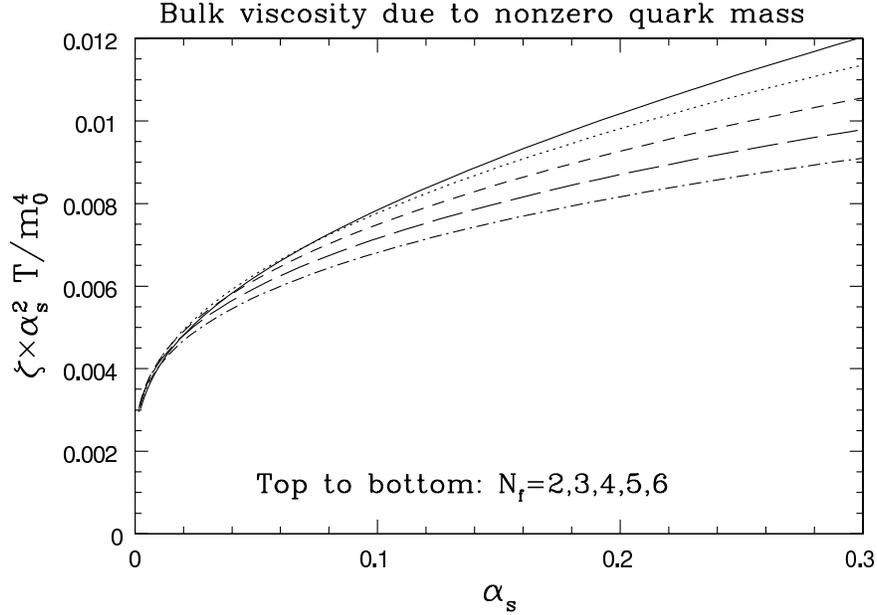}
\caption{\label{fig2}
Bulk viscosity when it is dominated by a single quark flavor's mass,
as a function of
$\alphas$, for several values of $\nf$.
}
\end{figure}

\begin{figure}
\centerbox{0.7}{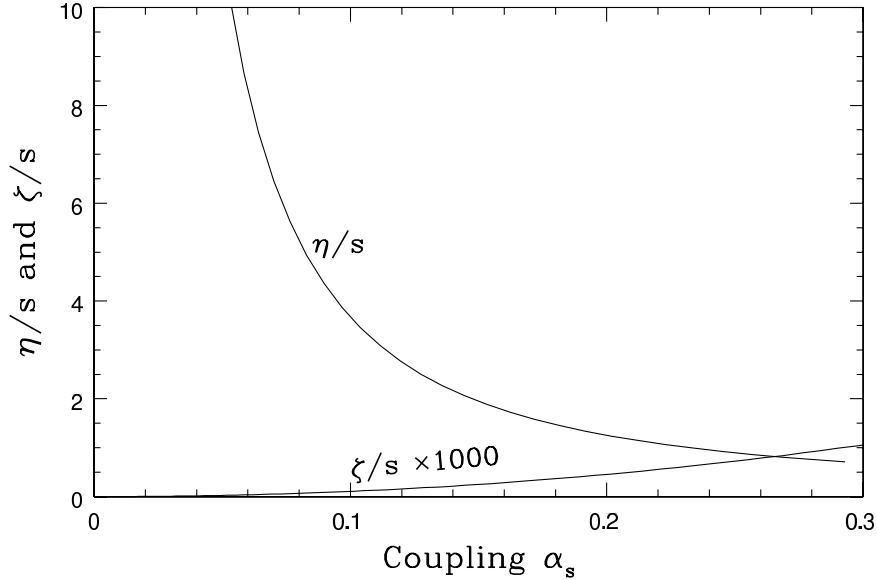}
\caption{\label{fig3}
Shear versus bulk viscosity:  $\eta/s$ and $\zeta/s$ ($s$ the entropy
density) as a function of $\alphas$, for $\nf{=}3$ QCD, neglecting quark
masses.  Bulk viscosity $\zeta$ has been rescaled by a factor of 1000.
}
\end{figure}

\begin{figure}
\centerbox{0.6}{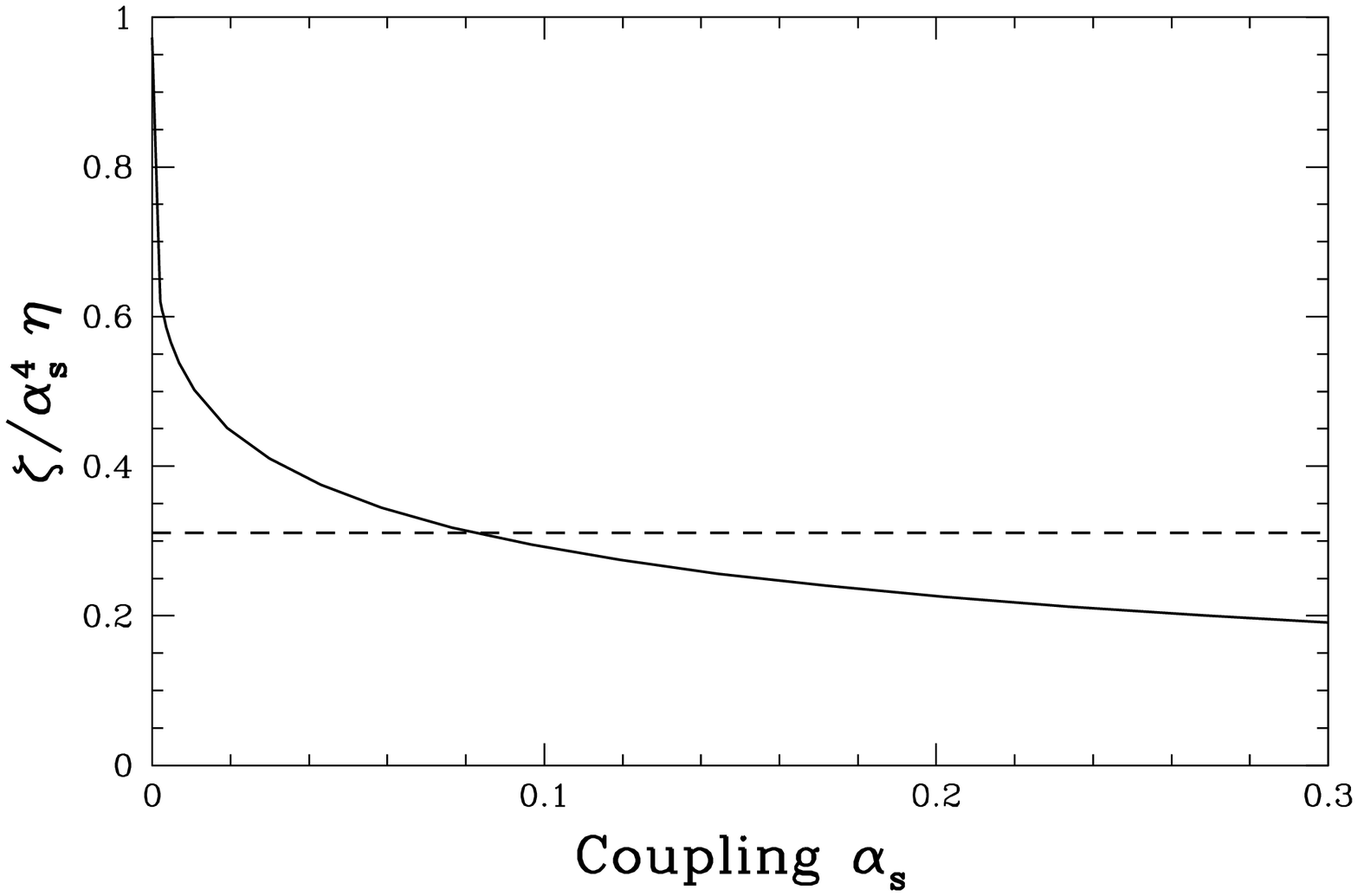}
\caption{\label{fig:ratio}
The ratio $\zeta/\alphas^4\eta$ for $\nf{=}3$ QCD, neglecting quark
masses.  The dashed line shows the crude estimate of
(\ref{eq:crudezeta}) with (\ref{eq:deltavs}).
As $\alphas\to0$ (and leading-log approximations to the leading-order
result become applicable), the ratio approaches the limit
$\zeta/\alphas^4\eta \to 0.973$.
}
\end{figure}

Section \ref{details} will present the details of the
calculation of bulk viscosity.
Our discussion will at times be abbreviated, referring back to
previous papers \cite{lead-log,lead-order}, where much of the technology
has already been presented.  We will end with a discussion in
section \ref{discussion}.  However,
for the impatient reader, we now present our main results.  The
coefficients, missing in \Eq{eq:param_result}, are presented in
Fig.~\ref{fig1} and Fig.~\ref{fig2}.
Here, $\nf$ is the number of flavors of quarks.
In Fig.~\ref{fig1}, all quark flavors are assumed to be massless
($m_0 \ll \alphas T$); in Fig.~\ref{fig2}, all but one flavor is
assumed to be massless, with that one flavor's mass in the range
$\alphas T \ll m_0 \ll \alphas^{1/2} T$.
A comparison of bulk viscosity and shear viscosity for three massless
flavors is given in Fig.~\ref{fig3} as a function of $\alphas$.
The figure makes clear that neglecting
bulk viscosity in favor of shear viscosity is actually quite a good
approximation, not only at weak coupling but probably also at
moderately strong, physically
interesting couplings.
Fig.\ \ref{fig:ratio} shows the ratio
$\zeta/\alphas^4\eta$, which at very small $\alphas$ approaches a constant
with corrections given by powers of $(\log(1/\alphas))^{-1}$.
The dashed line shows an old, crude estimate of the ratio of bulk
to shear viscosity which will be discussed in Sec.\ \ref{discussion}.

Throughout this paper, we will not attempt to project our leading-order
results to coupling higher than $\alphas \simeq 0.3$.  In previous studies
of diffusion constants \cite{lead-order}, it was found that this is where
different formulations of the effective kinetic theory, which were
equivalent at leading-order in coupling, no longer agreed within a
factor of 2, suggesting a complete breakdown of the perturbative
treatment.%
\footnote{
  See in particular Fig.\ 4 of Ref.\ \cite{lead-order} at
  $\md/T = 2.4$ for 3-flavor QCD, which corresponds to $\alphas = 0.3$.
}


\section{Physics of bulk viscosity}
\label{big_picture}

\subsection{Basic picture}
\label{sec:basic}

When a fluid is uniformly compressed, it leaves equilibrium.  The energy
density rises, but the pressure temporarily rises by more than what is
predicted by the equation of state.%
\footnote%
    {%
    That the pressure is higher during compression and lower during
    rarefaction is dictated by the second law of thermodynamics; if the
    pressure during compression were lower than in equilibrium, one
    could construct a perpetual motion machine of the second kind, which
    rapidly compressed a fluid (encountering a lower than thermodynamic
    pressure) and then slowly expanded a fluid (encountering the full
    thermodynamic pressure).  This constraint, that $\zeta$ is positive,
    is another way of seeing that $\zeta$ must be proportional to the {\em
    second} power of the beta function (or other source of conformal
    invariance breaking), since the beta function can be of either sign.
    }
Under uniform rarefaction, the
pressure temporarily falls further than is predicted by the fall in the
density and the equation of state.  The bulk viscosity quantifies the
time integral of this extra shift in the pressure (per e-folding of
expansion).

The change in pressure occurs because the fluid leaves equilibrium.  The
time scale for weakly coupled QCD to relax towards equilibrium is set by
the rate $\Gamma \sim \alphas^2 T \log[1/\alpha]$
for a typical particle ($p \sim T$) to randomize its
momentum $\p$.
The faster the fluid equilibrates, the nearer to equilibrium it remains,
so the smaller the shift in the pressure; therefore the viscosity should
be proportional to $\epsilon/\Gamma \sim T^3 / \alphas^2 \log[1/\alpha]$.
This naive estimate turns out to be parametrically correct for shear
viscosity.

However, it is wrong for bulk viscosity.  The reason is that QCD (at
high temperatures and away from mass thresholds) is a
nearly conformal theory, and the bulk viscosity vanishes in a conformal
theory, {\em for two reasons}.  First, uniform compression or
rarefaction is the same as a dilatation transformation.  In a conformal
theory, a dilatation transformation is a symmetry, and so the fluid will
not leave equilibrium.  Therefore, $\zeta$ must be proportional to the
breaking of conformal invariance.

Furthermore, in a conformal theory, even if the fluid {\em is} out of
equilibrium, the pressure still does not deviate from the value given by
the equation of state, which for a conformal theory is exactly $P =
\epsilon/3$.  This is just the tracelessness of the stress-energy tensor in a
conformal theory.  For instance, consider massless $\lambda \phi^4$ theory,
\begin{equation}
{\cal L} = \frac{1}{2} \partial_\mu \phi \partial^\mu \phi +
\frac{\lambda}{24} \phi^4 \, ,
\end{equation}
at finite $\lambda$ [with $({-}{+}{+}{+})$ metric convention].
The Euler-Lagrange equation is,
\begin{equation}
\partial^2 \phi - \frac{\lambda}{6} \phi^3 = 0 \, ,
\end{equation}
and the stress tensor and its trace are
\begin{equation}
T_{\mu\nu} = \partial_\mu \phi \partial_\nu \phi
- \eta_{\mu \nu} {\cal L} \, , \qquad
T^\mu_\mu = -(\partial \phi)^2 - \frac{\lambda}{6}\phi^4 \, .
\end{equation}
Multiplying the Euler-Lagrange equation by $\phi$ shows that $T^\mu_\mu$
vanishes up to a total derivative, which averages to zero.  This
argument is only flawed because of the conformal anomaly, which arises
because of the running of $\lambda$ with scale.
The bulk
viscosity coefficient will therefore contain another power of the smallness of
conformal invariance breaking

Thus, in a nearly conformal theory, the bulk viscosity coefficient
$\zeta$ vanishes as the {\em second} power of the departure from
conformality: one power because the departure from equilibrium is small,
and another power because any departure from equilibrium has a small
impact on the pressure.  For massless QCD, conformal symmetry is broken
by the running of the coupling, $\beta(\alphas) \sim \alphas^2$, and so
the bulk viscosity is
\begin{equation}
\zeta \sim \frac{T^3}{\alphas^2 \log[1/\alphas]} \times (\alphas^2)^2
\sim \frac{\alphas^2 T^3}{\log[1/\alphas]} \, ,
\end{equation}
as claimed before.  The presence of quark masses also constitutes a
breaking of conformal invariance provided $\mo \lsim T$ (otherwise there
are no quarks in the thermal bath and the influence of the quark can be
neglected).  In this case the pressure deviates from the massless value
by a relative amount $\sim \mo^2/T^2$, and
\begin{equation}
\zeta \sim \frac{T^3}{\alphas^2 \log[1/\alphas]} \times
\left(\frac{\mo^2}{T^2} \right)^2 \sim
\frac{\mo^4}{T \alphas^2 \log[1/\alphas]} \, .
\end{equation}

For future reference, note that if one formally defines the pressure
as $P = T_{ii}/3$ and linearizes the hydrodynamic formula (\ref{eq:Tij})
about global equilibrium $P=\Peq(\epsilon)$,
then the bulk viscosity parametrizes
\begin {equation}
   \Delta P - v_{\rm s}^2 \, \Delta\epsilon = \zeta \grad\cdot\uvec ,
\label {eq:DeltaP}
\end {equation}
where $v_{\rm s}$ is the velocity of sound, given by
$v_{\rm s}^2 = \partial \Peq  / \partial\epsilon$, and $\Delta P$
and $\Delta \epsilon$ are the local deviations of pressure and energy density.


\subsection{Number changing processes:
            Comparison with \boldmath$\phi^4$ theory}

There is one detail which this brief discussion has brushed over.
Viscosities are typically determined by the {\it slowest}\/ process
which is required for relaxation to equilibrium.
Certain departures from equilibrium can be very slow to equilibrate, due
to the presence of almost-conserved quantities.

\begin{figure}
\centerbox{0.4}{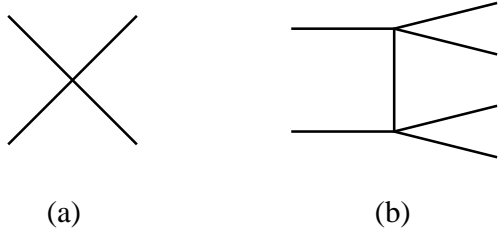}
\caption{\label{fig:scalar}
Examples of
(a) number conserving and (b) number changing
processes in $\phi^4$ theory.
}
\end{figure}

For instance, when
considering bulk viscosity in scalar $\lambda \phi^4$ theory, Jeon
found \cite{Jeon} that the total particle number equilibrates very
slowly.  The dominant process which randomizes momenta and determines
the shear viscosity is shown in Fig.\ \ref{fig:scalar}a, with rate
$\Gamma \sim \lambda^2 T$.  In contrast, an example process which
changes particle number, required for bulk viscosity, is shown in
Fig.\ \ref{fig:scalar}b.
One might naively expect that the particle number changing rate
from such processes is
$\Gamma \sim \lambda^4 T$, but this misses a soft
enhancement.  Number change primarily occurs between low energy
excitations, where Bose stimulation increases the rate.  The correct
estimate is that the number of excitations relaxes at a
rate $\Gamma \sim \lambda^3 T$, but this is still parametrically
small compared to the $2\to2$ scattering processes of Fig.\ \ref{fig:scalar}a.
This leads to the result
$\zeta \propto \lambda T^3$ in $\phi^4$ theory, up to logarithms
\cite{Jeon}.  In scalar theory, number-changing processes are the
bottleneck for the relaxation to equilibrium characterized by
bulk viscosity.

\begin{figure}
\centerbox{0.5}{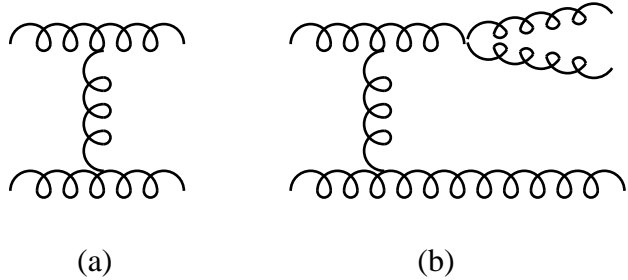}
\caption{\label{fig:brem}
Examples of (a) number conserving and (b) number changing
processes in QCD.
}
\end{figure}

The same does not occur in QCD (at vanishing chemical potential%
\footnote%
    {%
    For the Standard Model
    at finite baryon number chemical potential $\mu$ and finite
    quark mass $\mo$, the bulk viscosity would be very large.  Compressing
    the system changes the temperature, which shifts how much of the
    baryon number is stored in each quark type, in equilibrium.  The
    actual distribution of baryons between quark types approaches
    this equilibrium value only by weak interactions,
    leading to a bulk viscosity $\zeta \sim \mu^2 \mo^4/G_{\rm F}^2 T^7$
    for $\mu \lesssim T$ in cases where this is the rate-limiting
    process.
    Note however that in the early universe $\mu/T \sim 10^{-9}$ is
    negligible, while in a heavy ion collision weak interactions can be
    neglected entirely and one should take the numbers for each quark
    type to be separately conserved.
    }%
), however, because
number changing processes are much more efficient in gauge theory.%
\footnote{
  See, for example, Ref.\ \cite{sansra} and the related discussion of
  photon Bremsstrahlung in Ref.\ \cite{aurenche}.
}
The analog of Fig.\ \ref{fig:scalar} is Fig.\ \ref{fig:brem}.
Number change is relatively fast even among hard particles
and occurs
by $1{\leftrightarrow}2$ splitting of a hard particle into two other
hard particles during a
small-angle collision, such as depicted by Fig.\ \ref{fig:brem}b.
The small-angle collision rate is of order
$\alphas T$, and the nearly collinear emission from such scatterings costs
one extra factor of $\alphas$, giving a hard splitting rate
$\Gamma \sim \alphas^2 T$.
For comparison, the rate for a hard
particle to randomize its momentum through $2{\leftrightarrow}2$ collisions is
of order $\alphas^2 T \log[1/\alphas]$, which is larger by a
logarithm.  One might then suppose that number change is still the
bottleneck process for bulk viscosity (by a logarithm), that the
relevant rate is therefore $\alphas^2 T$ rather than
$\alphas^2 T \log[1/\alphas]$, and that therefore there should
be no logarithm in the parametric formula \Eq{eq:param_result}
for $\zeta$.
This turns out
not to be the case, though, because $2{\leftrightarrow}2$
scattering processes exchange gluons
between hard and soft momenta efficiently, and {\em soft} gluon number
changing processes are efficient enough to prevent a particle number
chemical potential from developing.
In section \ref{sec:zeromodes}, we will show that, because of
Bose stimulation enhancements for soft gluon emission from hard
particles, the
total rate of number-changing processes per particle
is $O(\alphas^{3/2} T)$, which
is parametrically faster than the $O(\alphas^2 T \log)$ rates discussed
above.  It is the latter, $O(\alphas^2 T \log)$
rates that will therefore be the bottleneck
for equilibration and which will determine the QCD bulk viscosity.

\section{Details of the calculation}
\label{details}

\subsection{Overview}

We now proceed with the details of the calculation of bulk viscosity.
Our general approach and notation will follow \cite{lead-log}.
To begin, note that, at weak coupling, there are long lived
quasiparticles, and a kinetic theory treatment should be valid.  The
plasma is well described by a phase space density for each particle type,
$f(\x,\p)$, which can be expanded about a local equilibrium distribution
$\feq(\x,\p)$ as
\begin{eqnarray}
f(\x,\p,t) & = & \feq(\x,\p,t) + f_1(\x,\p,t) \, , \nonumber \\
\feq(\x,\p,t) & = & \left( \exp \left[ 
\beta(t) \gamma_\u
     (E_\p - \p \cdot \uvec(\x))\right] \mp 1 \right)^{-1} \, ,
\end{eqnarray}
with $\beta \equiv T^{-1}$ and $\gamma_\u \equiv (1-u^2)^{-1/2}$.
The departure from equilibrium is
determined by the Boltzmann equation,
\begin{equation}
\frac{\partial f}{\partial t} + \v_\p \cdot \grad_{\x} f = - C[f] \, ,
\label{eq:boltzmann}
\end{equation}
with $C[f]$ the collision integral.  Above, $E_\p$ and
\begin {equation}
   \v_\p \equiv \grad_\p E_\p
\label {eq:vp}
\end {equation}
are the energy and velocity of a particle with momentum $\p$.%
\footnote{
  We use the general formula (\ref{eq:vp}) [which can be understood as the
  group velocity of a wave packet] because we would like to make a
  general treatment of quasi-particles with some dispersion relation
  $E_{\rm p}$, and there is no need at this point to specialize,
  for example, to $E_\p^2 = p^2 + m^2$.
}
To study transport coefficients such as viscosity, we are interested
in small departures from equilibrium in the hydrodynamic limit of
slow variation in $\x$ and $t$.
The left-hand side of (\ref{eq:boltzmann}) is explicitly small because of
the derivatives, and so we may replace $f$ by $\feq$ there.
The collision term must
be expanded to first order, $C[f] \propto f_1$,
noting that $C[\feq] = 0$ by local detailed balance.

It is convenient to analyze the problem in a local region, choosing
an approximate rest frame where
$\u(\x)$ and the variation of $\beta(t)$ can be taken to be small.
To first order in these small quantities, the derivatives appearing
on the left-hand side of the Boltzmann equation are
\begin {equation}
   \partial_\mu \feq(\x,\p,t)
   = - f_0(E_\p) [1 \pm f_0(E_\p)] \;
      \partial_\mu \bigl[ \beta(t) \, (E_\p - \p\cdot\uvec(\x)) \bigr]
      \biggl|_{\beta(t)=\beta,~\uvec = 0} ,
\label {eq:derivs}
\end {equation}
where $f_0$ is the Bose or Fermi distribution
\begin {equation}
   f_0(E) = (e^{\beta E} \mp 1)^{-1} .
\end {equation}
The departure from equilibrium, in the case of bulk viscosity, arises
because 
\begin{equation}
\grad \cdot \uvec \equiv X \neq 0 \, .
\end{equation}
In Sec.\ \ref{sec:q} below, we will use the derivatives
(\ref{eq:derivs}) and thermodynamic relations (in a treatment
slightly generalizing that of Jeon and Yaffe \cite{JeonYaffe}) to
rewrite the left-hand side of the Boltzmann equation (\ref{eq:boltzmann})
in the form
\begin{equation}
\frac{\partial\feq^a}{\partial t} + \v_\p \cdot \grad_{\x} \feq^a
= \beta f_0 (1{\pm}f_0)\, X(\x) \, q^a(\p)
\label {eq:qdef}
\end{equation}
for the case of isotropic compression or expansion,
relevant to bulk viscosity.
Here, $a$ is a species label, and
$q^a(\p)$ represents how much a particle of type $a$ and momentum $\p$
contributes to the $\Delta P - v_{\rm s}^2 \Delta\epsilon$
of (\ref{eq:DeltaP}).

The departure $f_1$ from local equilibrium, at linearized order, must also
be proportional to $X(\x)$, and it is convenient to parametrize it as
\begin{equation}
f^a_1(\x,\p) = \beta^2 f_0 (1{\pm} f_0) X(\x) \chi^a(|\p|) .
\end{equation}
The function $\chi(|\p|)$ will be a nontrivial
function of the magnitude of momentum $p\equiv |\p|$, but (in the local
rest frame) it is direction independent, because $X$ is a scalar
quantity.
Defining
\begin{equation}
{\cal S}^a(\p) = -T q^a(\p) f_0(1{\pm}f_0) \, ,
\label {eq:Sdef}
\end{equation}
the Boltzmann equation can be written as
\begin{equation}
  \S^a(\p) = [{\cal C}\chi]^a(\p) \, ,
\label{eq:Sboltz}
\end{equation}
with ${\cal C}$ the linearization of the collision integral, which
we will give in Sec.\ \ref{sec:C}.

The bulk viscosity is then determined as the shift in the pressure
induced by the departure from equilibrium $\chi$.
As we shall discuss,
this is an integral over $\p$ of $\chi^a$ times the same source $\S^a$ already
introduced,
\begin{equation}
\zeta
= \beta^3 \sum_a \nu_a \int \frac{d^3 \p}{(2\pi)^3} \, \S^a(p) \, \chi^a(p) 
 \equiv \Big( \S \, , \, \chi \Big) \, ,
\label{eq:innerprod}
\end{equation}
where $\nu_a$ is the multiplicity of species type $a$.
The collision operator ${\cal C}$ is Hermitian under this inner product,
and we may formally write,
\begin{equation}
\zeta = \Big( {\cal S} \, , \, {\cal C}^{-1} {\cal S} \Big) \, .
\label{eq:zeta}
\end{equation}
This can then be treated variationally, by the techniques
presented in Refs.\ \cite{lead-log,lead-order}.


\subsection{General formula for \boldmath$q^a(p)$}
\label{sec:q}

It remains to determine $q^a$, to establish the form of ${\cal C}$, and
to explain how the integral equations will be solved.  We will treat
$q^a$ first, since it is the most different from the problems already
addressed in Refs.\ \cite{lead-log,lead-order}.
The second term on the lefthand side 
of \Eq{eq:boltzmann} is
\begin{equation}
\v_\p \cdot \grad_{\x} f = 
 \beta f_0(1{\pm} f_0)\; p_i v_{\p,j} \nabla_j \u_i
\label{eq:source1}
\end{equation}
at linearized order.  Specializing to isotropic compression or
expansion, $\nabla_i u_j = (\delta_{ij}/3) \grad\cdot\uvec$, this
becomes
\begin{equation}
\v_\p \cdot \grad_{\x} f = \beta f_0 (1{\pm} f_0) \;
\grad \cdot \uvec \; \frac{\v_\p \cdot \p}{3} \, .
\label{eq:source2}
\end{equation}

Unlike the case of shear viscosity, the term $\partial_t f$ is also
nonzero; the compression or rarefaction of the fluid causes its density,
and therefore its temperature, to change with time.  By the chain rule,
\begin{equation}
\frac{\partial f_0}{\partial t}
= \frac{d\beta}{dt} \frac{\partial f_0}{\partial\beta}
= - \frac{d\beta}{dt} f_0(1{\pm}f_0) \frac{\partial(\beta E_\p)}{\partial\beta}
  \, .
\label {eq:sourcedt}
\end{equation}
Now,
stress-energy conservation implies
\begin{equation}
\partial_\mu T^{\mu \nu} = 0 \qquad \rightarrow \qquad
\partial_t \epsilon = - (\epsilon + P) \grad \cdot \uvec \, .
\label {eq:conserve1}
\end{equation}
By standard thermodynamic relations,
\begin{equation}
\epsilon + P = T \frac{dP}{dT}
\end{equation}
(recall that $P=-F$ for a theory without chemical potentials), and by
the chain rule,
\begin{equation}
\frac{d\epsilon}{dt} = \frac{dP}{dt} \frac{d\epsilon}{dP}
= v_{\rm s}^{-2} \partial_t P \, .
\label {eq:conserve3}
\end{equation}
Combining (\ref{eq:conserve1}) thru (\ref{eq:conserve3}),
\begin{equation}
\frac{dP}{dt} =  - v_{\rm s}^2 \grad \cdot \uvec T \: \frac{dP}{dT}
= v_{\rm s}^2 \grad \cdot \uvec \: \beta \frac{dP}{d\beta}
\, ,
\end{equation}
and since the dependence of $P$ on $t$ is through its $\beta$
dependence, it follows that
\begin{equation}
\frac{d\beta}{dt} = \beta v_{\rm s}^2 \: \grad \cdot \uvec \, .
\label{eq:dbdt}
\end{equation}
Therefore, combining (\ref{eq:source2}), (\ref{eq:sourcedt}),
and (\ref{eq:dbdt}),
the full lefthand side of the Boltzmann equation is
\begin{equation}
\left( \partial_t + \v_\p \cdot \grad_\x \right) f_0 = 
\beta f_0 (1{\pm} f_0) \grad \cdot \uvec \left(
\frac{\p \cdot \v_\p}{3}
  - v_{\rm s}^2 \frac{\partial(\beta E_\p)}{\partial\beta}
\right) \, .
\label{eq:source3}
\end{equation}
Comparing to the definition (\ref{eq:qdef}) of $q^a(\p)$, we determine
\begin{equation}
q^a(\p) = \frac{\p \cdot \v_\p}{3} - v_{\rm s}^2
\frac{\partial(\beta E_\p)}{\partial\beta} \, .
\label{eq:q}
\end{equation}

A nice property of this formula is that one can easily verify that the
source vanishes in a conformal theory.  In a conformal
theory, the only dimensionful scale would be $T$, and so, by dimensional
analysis, $E_\p$ must
have the form $E_\p = p \, F(p/T)$ for some
function $F$.  Using (\ref{eq:vp}), (\ref{eq:q}), and the conformal
result $v_{\rm s}^2 = \third$, one would
then find $q^a(\p) = 0$.

Before finding explicit expressions for $E_\p$ and $v_{\rm s}^2$ in QCD,
let us briefly discuss the result for $q^a(\p)$.
The $\p\cdot \v_\p$ term represents the
change $v_\p\cdot\grad_\x \feq$
in the quasiparticle distribution function due to free
propagation.  For all other transport coefficients we have computed
\cite{lead-log,lead-order}, this type of change
was the appropriate ``source'' in the
Boltzmann equation, and the collision integral was to be equated with
it.  But here the ``source'' has nonvanishing energy, and energy is
conserved.  The collision integral has an exact zero mode associated
with energy conservation; therefore collisions will not erase the change
in $f$, but will re-distribute it until it looks like a shift in the
temperature.  The size of the temperature shift is fixed by energy
conservation---that is, by the amount of energy the $\p\cdot \v_\p$ term
introduces.  Therefore, the true departure
from equilibrium is the difference between this $\p\cdot \v_\p$ source
term, and the temperature shift which carries the same total energy.
This is the role of the second
$v_{\rm s}^2 \partial(\beta E_\p)/\partial\beta$ term.
In other words, considering the linearized collision operator ${\cal C}$
as an operator on the space of departures from equilibrium $\delta f$,
we must project the source $\p \cdot \v_\p$ into the subspace orthogonal to
the zero mode of ${\cal C}$ (since the eigenvector of the zero mode is
not actually a departure from equilibrium).

As a check, we give a general
demonstration in Appendix \ref{app:orthogonal}
that the source term determined by (\ref{eq:q})
indeed carries no energy in the quasi-particle approximation we have
used throughout.  (One may also eschew generality and instead
directly check with the explicit formulas for $q^a(\p)$ given
in the next section.)  In the appendix, we also discuss in more detail why
the quasi-particle approximation is justified for a leading-order
calculation of the bulk viscosity.

With this in mind, we can see why it is this same $q^a(p)$ which is
relevant in determining the pressure shift due to the departure from
equilibrium $\propto \chi^a(p)$.  Naively, the extra pressure due to
a departure from equilibrium $f_1(\p)$ should be $\frac 13 \int_\p f^a_1(\p) \;
\p\cdot \v_\p$.  However, a general shift in the equilibrium
distribution function by $f_1$ leads to a shift in the energy.
Bulk viscosity involves the difference between the actual pressure, and
the pressure determined by $\epsilon$ and thermodynamics, $P(\epsilon)$.
Therefore, we must subtract off $(dP/d\epsilon)\delta \epsilon
=v_{\rm s}^2 \delta \epsilon$, the shift in the pressure due to the
extra energy density contributed by $f_1$.  That is precisely what the
second term in \Eq{eq:q} does.%
\footnote{
   This subtraction is technically unnecessary if one has
   already projected the source to be orthogonal to the zero mode, since
   then no shift in the energy would be produced.
   However, it is convenient, because it allows a symmetric treatment
   of the source and the pressure shift, as is manifested by the
   symmetric appearance of $S$ in (\ref{eq:zeta}).
}


\subsection{Specific formula for \boldmath$q^a(p)$}
\label{sec:qspecific}

Now we will determine in detail the form $q^a(\p)$ takes in QCD at weak
coupling $\alphas \ll 1$.  For simplicity we will also take quark masses
$\mo \ll T$, though nothing in principle stops us from considering the
case of quark masses $\mo \sim T$.
We will assume that there is at most one quark
species with non-negligible quark mass, which we denote $\Mo$.
In this case, the energy of a quasiparticle excitation of
momentum $p \gg gT$, to first order in $g^2$, is given by
\begin{eqnarray}
E_\p^2 & = & p^2 + m_{\infty}^2 \, , \nonumber \\
m_{\infty,a}^2[\mbox{quark}] & = & \moa^2 + \frac{\cf g^2 T^2}{4} 
\hspace{0.38in}= \moa^2 + \frac{g^2 T^2}{3} \, ,
\nonumber \\
m_{\infty}^2[\mbox{gluon}] & = & \left( \ca + \nf \tf \right)
\frac{g^2 T^2}{6} 
\hspace{0.1in}
= \frac{6 + \nf}{12} g^2 T^2 \, ,
\end{eqnarray}
where $\moa$ is the mass of quark species $a$.
The masses $m_\infty$ here are the corrections to the large $p$ dispersion
relations.%
\footnote{
  Their relation to frequently-used zero-momentum masses are
  $m_\infty^2 = m_{\rm D}^2/2 = 3 m_{\rm pl}^2/2$ for gluons and
  $m_\infty^2 = 2 m_{\rm F}^2$ for massless quarks, where
  $m_{\rm D}$ is the Debye mass, $m_{\rm F}$ is the analogous
  screening mass for quark exchange, and $m_{\rm pl}$ is the plasma
  frequency.  For further details, see, for example, Refs.
  \cite{Thoma,boltzmann}.
}
We have written these expressions in terms of group Casimirs so that
they can be evaluated for a general group, and have also given the
specialization to QCD, where the adjoint Casimir $\ca=3$, and the
fermions are in a representation with Casimir $\cf=4/3$ and trace
normalization $\tf=1/2$.  Here $\nf$ is the number of light Dirac
fermions, or half the number of Weyl fermions.

Using these expressions, $q^a$ becomes
\begin{eqnarray}
\label{eq:source4}
q^a & = & 
\frac{1}{E_\p} \left( \frac{\p^2}{3} - v_{\rm s}^2 (\p^2+\tilde m_a^2)
\right) \, , \\
\tilde m_a^2 & \equiv &
m_{\infty,a}^2 - \frac{d(m_{\infty,a}^2)}{d(\ln T^2)} 
\, ,
\label{eq:mtilde}
\end{eqnarray}
which coincides with the results of Jeon and Yaffe \cite{JeonYaffe}.

The speed of sound can be determined by writing out the temperature
dependence of the pressure.
At order $g^2$ and $\Mo^2$, the pressure of
the QCD plasma is \cite{Kapusta}
\begin{eqnarray}
\label{eq:P_is}
P &=& \Big(a + b g^2[\mu^2=T^2]\Big)T^4 + c \Mo^2 T^2 \, , \nonumber \\
a & = & \frac{\pi^2}{180}\left( 4\da + 7\nf \df \right) 
= \frac{\pi^2}{180}\left(32+21\nf \right) \, ,
\nonumber \\
b & = & \frac{-1}{288}\left( 2\da \ca + 5 \nf \df \cf \right)
= \frac{-1}{288} \left( 48 + 20 \nf \right) \, ,
\nonumber \\
c & = & \frac{-1}{12} \df = \frac{-1}{4} \, ,
\end{eqnarray}
where $\da=8$ and $\df=3$ are the dimensions of the adjoint
and fermion color representations.
Using $\epsilon=TdP/dT - P$, one finds
\begin{equation}
v_{\rm s}^2 = \frac{dP}{d\epsilon} = \frac{dP/dT}{d\epsilon/dT}
= \frac{1}{3} - \frac{2b}{9a} \beta(g^2) + \frac{c\Mo^2}{9aT^2} \, ,
\end{equation}
up to $O(g^5)$, $O(\mo^2 g^2/T^2)$, and $O(\mo^4/T^2)$ corrections.
Here,
\begin{equation}
\beta(g^2) \equiv \frac{\mu^2 dg^2}{d[\mu^2]} = \frac{g^4}{16\pi^2}
\left( \frac{4\nf \tf - 11 \ca}{3} \right)
= \frac{g^4}{16\pi^2} \left( \frac{2\nf - 33}{3} \right)
\end{equation}
is the beta function of QCD, which enters on taking the temperature
dependence of $g^2$ into account%
\footnote%
    {%
    We are implicitly taking $Td/dT$ holding $\mu/T$ fixed.  But we
    would get the same answer if we performed the derivative holding
    $\mu$ fixed; in writing $g^2[\mu^2=T^2]$ in
    \Eq{eq:P_is}, what we really mean is that there is explicit $\mu$
    dependence in the $g^4$ term, of form 
    $B \beta(g^2) \log(T^2/\mu^2) T^4$.  Holding $\mu$ fixed,
    $\beta(g^2)$ arises from the $T$ derivative of this logarithm.
    }.
Similarly, the quantities $\tilde m^2$ introduced earlier involve $\mo^2$
and $\beta(g^2)$, and are
\begin{eqnarray}
\tilde m^2[\mbox{quark}] & = & \moa^2 - \frac{\cf T^2}{4} \beta(g^2) \, ,
\nonumber \\
\tilde m^2[\mbox{gluon}] & = & -\frac{(\ca + \nf \tf)T^2}{6}\beta(g^2)
\, .
\end{eqnarray}
Collecting these results, and making the approximation
\begin{equation}
q^a = \frac{1}{E_\p} \left( \frac{\p^2}{3} - v_{\rm s}^2(\p^2+\tilde
m_a^2) \right) 
\simeq \frac{1}{p} \left( (\third-v_{\rm s}^2) \p^2
- \third \tilde m_a^2 \right) \, ,
\end{equation}
valid for $\mo^2/T^2 \ll 1$ and $\beta(g^2) \ll 1$, we find
\begin {subequations}
\label{eq:q_is}
\begin{eqnarray}
  q^{q,a} & = & |\Delta v_{\rm s}^2| \, p +
\left[ \frac{\cf}{12} \beta(g^2) \, T^2 -\frac{\moa^2}{3} \right] p^{-1}
\, ,
\label {eq:qq}
\\
q^g & = & |\Delta v_{\rm s}^2| \, p +
\left[ \frac{\ca \da + \nf \df \cf}{18\da}\beta(g^2) \, T^2
\right] p^{-1} \, , \qquad
\end{eqnarray}
\end {subequations}
where
\begin {equation}
| \Delta v_{\rm s}^2 | =
\frac{-5(2\da\ca + 5\df\cf\nf)
\beta(g^2) + 60 \df \Mo^2/T^2}{36\pi^2(4\da+7\df\nf)} \,.
\label {eq:deltavs}
\end {equation}
Here, the $\Mo^2$ in (\ref{eq:deltavs}) appears in the $q$ for
every species, but the
$\moa^2$ in the second term of (\ref{eq:qq})
only contributes to the possibly
massive quark, for which $\moa=\Mo$.
Note that, as promised, $q^a$ is proportional to the source of conformal
invariance violation, either the beta function or the current quark
mass.  Because ${\cal S}\propto q$ enters quadratically in
\Eq{eq:zeta}, we see that $\zeta$ will depend quadratically on the size
of conformal invariance breaking, as claimed.


\subsection{Variational method and the collision integral}
\label{sec:C}

It remains to specify the form of the collision integral, and to explain
how it will be inverted to establish $\zeta$ using \Eq{eq:zeta}.  Since
the details here are rather similar to the previous literature
\cite{lead-log,lead-order}, we will be somewhat brief in our
discussion.  First, define an inner product as in \Eq{eq:innerprod}
(summation over species label and integration over momenta).  Then the
Boltzmann equation and bulk viscosity can be formulated variationally;
define 
\begin{equation}
Q(\chi) \equiv \Big( \chi \, , \, \S \Big) - \frac{1}{2}
\Big( \chi \, , \, {\cal C} \chi \Big) \, ,
\end{equation}
and observe that $\delta Q/\delta \chi = 0$ when $\chi$ satisfies the
Boltzmann equation (\ref{eq:Sboltz}).
Furthermore, the value (\ref{eq:zeta}) of $\zeta$ is $2 Q$
evaluated at this extremum:
\begin {equation}
   \zeta = 2 Q_{\rm max} .
\label{eq:zetaQmax}
\end {equation}
A variational {\it Ansatz} for $\chi$ will
give a lower bound on the value of the extremum which will improve
rapidly as the variational basis is increased.  Therefore, we 
write a multi-parameter, linear {\it Ansatz}
for $\chi(p)$, in terms of a set of basis functions.
As we will discuss momentarily, $\chi(p) \propto p$ at small momenta,
and $\chi$ grows no faster than $\sim p^2/T$ at large $p$ .  Therefore,
we use a slight modification of the basis functions considered in
\cite{lead-log},
\begin{equation}
\label{eq:basis}
\phi_m(p) = \frac{p^{m} T^{K-m-1}}{(T+p)^{K-2}} \, , \qquad
m=1\ldots K \, .
\end{equation}
The function $\chi^a(p)$ is then assumed to be of form,
\begin{equation}
\chi^a(p) = \sum_m \tilde\chi^a_m \phi_m(p) \, .
\end{equation}
Within this variational {\it Ansatz}, the required inner products for
$Q$ are
\begin{eqnarray}
\Big( \chi \, , \, \S \Big) & = & \sum_{a,m} \tilde\chi^a_m
\tilde S^a_m \, , \nonumber \\
\Big( \chi \, , \, {\cal C} \chi \Big) & = & \sum_{abmn}
\tilde\chi^a_m \tilde{C}^{ab}_{mn} \tilde\chi_n^b \, ,
\\
\noalign {\hbox{where}}
\tilde{S}^a_m & \equiv & \nu_a \int_\p \phi_m(p) \, \S^a(p) \, ,
\nonumber \\
\tilde{C}^{ab}_{mn} & \equiv &
\nu_a \int_\p \phi_m(p) \, [{\cal C}^{ab} \phi_n](p) \, ,
\label {eq:Cab}
\end{eqnarray}
where ${\cal C}^{ab}$ means the collision integral for species $a$ when
species $b$ is out of equilibrium by the amount indicated by $\chi^b$.
Considering
$\tilde S^a_m$ to be a rank $NK$ column vector $\tilde S$ and
$\tilde{C}^{ab}_{mn}$ to be a $NK\times NK$ matrix, where $N$ is
the number of possibilities for the species index $a$,
the bulk viscosity (\ref{eq:zeta}) is
\begin{equation}
\zeta = \tilde S \tilde C^{-1} \tilde S \, .
\label{eq:C_inv}
\end{equation}
In practice, $N=2$ (quarks vs.\ gluons) if all quarks are massless,
and $N=3$ (massive quark vs.\ massless quarks vs.\ gluons) if one quark
species is massive.

The detailed form of ${\tilde C}$ is given in Ref.\ \cite{lead-order},%
\footnote{
  See specifically Eqs. (2.22) and (2.23) of Ref.\ \cite{lead-order}
  with $\chi^a_{i..j}$ replaced by $\chi^a(p)$ to specialize to the
  isotropic ($l=0$) angular dependence relevant to bulk viscosity, and
  then define $\tilde C^{ab}_{mn}$ as in (\ref{eq:Cab}) of this paper.
}
which we summarize here for completeness:
\begin{eqnarray}
\label{Collision_int}
\tilde C^{ab}_{mn} & \equiv & \frac{\beta^3}{8} \sum_{cdef}
\int_{\p\k\p'\k'} |{\cal M}^{cd}_{ef}(\p,\k;\p',\k')|^2
(2\pi)^4 \delta^{(4)}(P{+}K{-}P'{-}K')
\nonumber \\ &&\hspace{1in}\times
f_0^c(p)f_0^d(k)[1{\pm}f_0^e(p')][1{\pm} f_0^f(k')]
\nonumber \\ &&\hspace{1in}\times
\left[
\phi_m(p) \delta^{ac} {+} 
\phi_m(k) \delta^{ad} {-} 
\phi_m(p')\delta^{ae} {-} 
\phi_m(k')\delta^{af}\right]
\nonumber \\ &&\hspace{1in}\times
\left[
\phi_n(p) \delta^{bc} {+} 
\phi_n(k) \delta^{bd} {-} 
\phi_n(p')\delta^{be} {-} 
\phi_n(k')\delta^{bf}\right]
\nonumber \\ && +
\frac{\beta^3}{2} \sum_{cde} 4\pi
\int_0^\infty dp'\, dp\, dk \> \gamma^c_{de}(p';p,k) \delta(p'{-}p{-}k)
f_0^c(p')[1{\pm}f_0^d(p)][1{\pm} f_0^e(k)]
\nonumber \\ &&\hspace{0.5in}\times
\left[
\phi_m(p')\delta^{ac} {-} 
\phi_m(p) \delta^{ad} {-} 
\phi_m(k) \delta^{ae}\right]
\left[
\phi_n(p')\delta^{bc} {-} 
\phi_n(p) \delta^{bd} {-} 
\phi_n(k) \delta^{be}\right] . \qquad
\end{eqnarray}
All factors of the number of degrees of freedom of each species
are implicitly included in these sums.%
\footnote{
  In the convention of Ref.\ \cite{lead-order}, the sums (no averages)
  over all initial and final colors are included in $|{\cal M}|^2$ and
  $\gamma$, and each of the indices $cdef$ in the explicit sums above
  denote gluons vs.\ different flavors of quarks vs.\ different flavors
  of anti-quarks.
}
The detailed expressions for the $2{\leftrightarrow}2$ amplitude
${\cal M}$ and the effective $1{\leftrightarrow}2$ splitting
function $\gamma$ fill two
appendices of Ref.\ \cite{lead-order} and will not be reproduced here.
In treating the kinematics of these processes, we have neglected the
masses of all external states, which is consistent with our
approximation, $\mo \ll T$.  In principle there is no obstacle to
treating the case $\mo \sim T$, but we have not done so, primarily out of
laziness. 

Besides the difference in the source, which we have already stressed,
the other difference between bulk and shear viscosity calculations is in
the angular dependence of $\chi$ in the collision integral.  For shear
viscosity, it was not $\phi_m(p)$ which appeared above, but
$\phi_m(p) \, \hat{p}_i \hat{p}_j$.
(See, for example, Ref.\ \cite{lead-order}
for a discussion in the conventions of this paper.)  When suitably
averaged over the indices $ij$, this led to angular factors of
$P_2(\cos \theta_{\p\k})$
in the cross-term between $\chi_m(p)$ and $\chi_n(k)$,
for instance, where $P_2(x)=(3x^2{-}1)/2$ is the second Legendre
polynomial.  Since bulk viscosity arises due to $X=\grad \cdot \uvec$, a
scalar quantity, this angular dependence is absent.  This makes the
calculation of the collision integral somewhat simpler, but it does add
two complications involving zero modes of the collision operator, to
which we now turn.


\subsection{Zero modes of \boldmath${\cal C}$}
\label{sec:zeromodes}

The first term in the collision integral (\ref{Collision_int}), corresponding
to $2 \leftrightarrow 2$ processes, has two exact zero modes,
corresponding to all $\chi^a(p)\propto 1$ and all $\chi^a(p) \propto p$,
corresponding to particle number conservation and energy conservation,
respectively.  The second term, corresponding to collinear
$1\leftrightarrow 2$ processes, breaks particle number but still has the
zero mode corresponding to energy conservation.  Therefore, the
collision matrix $\tilde C$ will have a zero mode, and can potentially
have a second approximate zero mode to the extent that the
$2\leftrightarrow 2$ term is larger than the $1\leftrightarrow 2$ term.
Since the collision integral must be inverted in evaluating
\Eq{eq:C_inv}, we must address the exact zero mode.
We will see that making a leading-log expansion of bulk viscosity
(if such is desired) requires treating the $2\leftrightarrow 2$ term as
larger, by a logarithm, than the $1\leftrightarrow2$ term for $p \sim T$.
In order to understand why number-changing processes are not
a bottleneck for equilibration, and to understand expansions in
$[\log(1/\alphas)]^{-1}$,
we will need to address the approximate zero mode as well.
Both of these zero modes are specific to the case of isotropic
$\chi(p)$, and neither is relevant to the analysis of other
standard transport coefficients such as shear viscosity and flavor
diffusion constants.

The presence of an exact zero mode in the collision integral is not
problematic, precisely because the source ${\cal S}$ carries precisely
zero energy, and so is orthogonal to the zero eigenvector.  Therefore,
our previous expressions should be understood as valid in the subspace
orthogonal to the zero mode of ${\cal C}$.  In practice our basis of
functions $\phi_m^a$ are not restricted to this orthogonal subspace.
But the collision integral can be rendered invertible without changing
its behavior in the orthogonal subspace by adding a constant times the
projection operator for the pure temperature fluctuation (the zero
mode);
\begin{equation}
\tilde C^{ab}_{mn} \rightarrow \tilde C^{ab}_{mn}
+ \lambda
\left( \nu_a \int \frac{d^3p}{(2\pi)^3}\,E_\p \phi_m(p)
       f_0^a(1\pm f_0^a) \right)
\left( \nu_b \int \frac{d^3k}{(2\pi)^3}\,E_\k \phi_n(k)
       f_0^b(1\pm f_0^b) \right)
 \, ,
\end{equation}
for any positive $\lambda$.  This renders $\tilde{C}$ invertible; and
while $\tilde C^{-1}$ is $\lambda$ dependent, $\tilde C^{-1} \tilde S$
is not, since $\tilde S$ has zero projection onto the modified
direction.  In our numerical evaluations we have checked explicitly that
the determined value of $\zeta$ has no sensitivity to the added value of
$\lambda$.

Next, consider the possible approximate zero mode, $\chi^a(p)$ a
constant, corresponding to a chemical potential for particle number.%
\footnote{
  By ``particle number,'' we mean the sum of quark, anti-quark, and
  gluon numbers, not a difference like quark minus anti-quark
  number.
}
First note that the constant value must be the same for fermionic and
bosonic species, because the set of $2\leftrightarrow 2$ processes
includes fermionic pair annihilation to gluons, which contributes at
leading logarithmic order.  However, no
elastic $2\leftrightarrow 2$ scattering processes will drive a common
chemical potential for both quark and gluon number to zero.
For the case of bulk viscosity in $\lambda
\phi^4$ theory it was found that this played a major role in setting the
bulk viscosity \cite{JeonYaffe}.  In that theory,
$\chi \propto 1$ is an approximate zero mode of the full collision
operator: $(\chi|{\cal C}\chi)$ for $\chi = 1$
is parametrically small compared to typical
hard collision rates.%
\footnote{
  This permits a simplification in $\phi^4$ theory whereby one can
  avoid solving an integral equation and instead determine the leading-order
  bulk viscosity from a simple expectation value $(\chi|{\cal C}\chi)$
  for $\chi \propto 1$ \cite{JeonYaffe}.
}

However, for the bulk viscosity of QCD, this would-be zero mode actually
plays no role: the expectation $(\chi|{\cal C}\chi)$
for $\chi = 1$ is
parametrically large
rather than small compared to typical hard
scattering rates.
The reason is that, while number changing collinear
processes (the second term in \Eq{Collision_int}) are subdominant to
$2\leftrightarrow 2$ processes at generic momenta, they are very fast
at producing and destroying soft gluons.  To see this, let us
estimate the total rate for a hard particle to produce a soft
gluon of momentum $k$ by Bremsstrahlung.  Combine
(i) the $O(g^2 T)$ rate for small-angle scattering,
as in Fig.\ \ref{fig:brem}a, times (ii) a factor of $g^2$ for
absorbing or emitting the additional gluon in Fig.\ \ref{fig:brem}b,
times (iii) an initial or final state factor of $f(k)$ or
$1 \pm f(k)$ for that gluon, and (iv) a momentum integral
$dk/k$ (responsible for the logarithmically large rate of soft
bremsstrahlung emission in vacuum%
\footnote{
  In vacuum, there is an additional logarithmic factor for
  bremsstrahlung from an ultra-relativistic particle---a
  collinear logarithm $\int d^2 k_\perp/ k_\perp^2 \sim \ln(q/m)$,
  where $q$ is the momentum-transfer in the underlying $2 \to 2$
  collision.  In our case, the most frequent collisions are the
  small-angle ones, whose impact parameter is limited by Debye
  screening, and $q \sim m \sim gT$ so that there is no collinear log
  enhancement.
}%
).
$f(k) \sim T/k$ for small $k$, and
the result for the number changing rate $\Gamma_{1 \to 2}^{\rm total}$
is then%
\footnote{
  For $k \ll T$, the Landau-Pomeranchuk-Migdal (LPM) effect plays no
  role in gluon emission, as discussed qualitatively in Sec.\ 5.2 of
  Ref.\ \cite{boltzmann}.
  This is different
  than the case of soft {\it photon}\/ emission due to the $O(gT)$ thermal
  mass and scattering of the emitted gluon, either of which, for
  $k \ll T$, causes loss of the multiple-collision coherence
  that produces the LPM effect.
  Here is a quick argument: For small $k$, the internal hard particle
  line in Fig.\ \ref{fig:brem}b is off-shell in energy by an amount of order
  $\Delta E = E_{\p+\k} - E_\p - E_\k \sim (m_{\rm g}^2 + k_\perp^2)/(2k)$.
  The formation time of the gluon is therefore of order
  $(\Delta E)^{-1} \lesssim k / m_{\rm g}^2 \sim k / (g^2 T^2)$, which is
  small compared to the time $1/(g^2 T)$ between collisions when $k \ll T$.
}
\begin {equation}
  \Gamma_{1\leftrightarrow2}^{\rm total} \sim
  g^4 T \int \frac{dk}{k} f(k)
  \sim
  g^4 T^2 \int \frac{dk}{k^2}
  \,.
\label {eq:12total}
\end {equation}
The infrared divergence of the integral will be cut off by the effective
thermal mass $m \sim gT$ of the emitted gluon, so that
\begin {equation}
  \Gamma_{1\leftrightarrow2}^{\rm total} \sim
  \frac{g^4 T^2}{m}
  \sim 
  g^3 T \, .
\end {equation}
As discussed earlier,
this is parametrically faster than the $O(g^4 T \log)$ rate
to redistribute momenta between soft and hard particles, which is
the bottleneck which determines bulk viscosity.
The total rate $\Gamma_{1\leftrightarrow2}^{\rm total}$
for creating or absorbing
soft particles can therefore be taken as formally infinite for the
purpose of a leading-order calculation of bulk viscosity.

The same result can also be obtained, with some difficulty,
from \Eq{Collision_int}
of this paper
together with eqs.\ (B1-B6) of Ref.\ \cite{lead-order}, which
determine the splitting functions $\gamma$.  In particular, the
$k \ll p$ behavior of $\gamma^{\rm g}_{\rm gg}(p';pk)$ and
$\gamma^{\rm q}_{\rm qg}(p';pk)$ is $\gamma \sim (g^4 T) p^2/k$.
If we substitute $\phi_m = \phi_n = \chi = 1$
into the $1\leftrightarrow 2$ term in
\Eq{Collision_int}, the $k$ integration for $p \sim T$ then gives
the linear divergence $\int dk /k^2$ of (\ref{eq:12total}).

This means that a chemical potential is actually very
rapidly thermalized by number changing processes.  Any $\chi(k)$ which
falls more weakly than $\chi(k) \propto k$ at small $k$ will lead to a
divergent collision rate, meaning that such departures from equilibrium
are so efficiently equilibrated that we need not consider them.
Therefore we should restrict our {\it Ansatz} for $\chi$ to only
functions which are linear or higher powers of $k$ in the soft
region.  This justifies our choice in \Eq{eq:basis}.  Within this
subspace of functions $\chi$,
the $2\leftrightarrow 2$ part of the collision integral has
only one zero mode, that associated with energy conservation, which we
have already discussed.  Therefore the small $\alphas$ behavior will
indeed be $\zeta \propto \alphas^2 T^3/\log[1/\alphas]$, and one can
perform an expansion in logarithms of the coupling if desired.


\subsection{Expansion in \boldmath$\log[1/\alphas]$}

In Ref.\ \cite{lead-order} it was shown that an expansion in inverse powers of
$\ln[1/\alphas]$ works surprisingly well at small values of $\alphas$,
if it is carried to next-to-leading order.  As we have just seen, there
is no obstacle to making a similar expansion here.  We have done so, 
by following the procedure described in detail in Ref.\
\cite{lead-order}, but
we find that the expansion works much less well than in the case of
shear viscosity and number diffusion.  The reason is that the dominant
physics in shear viscosity and number diffusion is {\em angle change}.
The charge $q^a$ in that case is $1$ or $|\p|$ times a nontrivial
function of angle.  The departure from equilibrium, $\chi(\p)$, has
nontrivial angular dependence, but turns out to have very simple $|\p|$
dependence, so a one parameter {\it Ansatz} works very well.  In a
next-to-leading log treatment, one fixes the $|\p|$ dependence of
$\chi(\p)$ using the leading-log part of the $2\leftrightarrow 2$
processes and evaluates the collision integral using this fixed form of
$\chi(\p)$.  This works because this functional form of $\chi(\p)$ is
essentially correct, whatever collision processes are involved.

\begin{figure}
\centerbox{0.7}{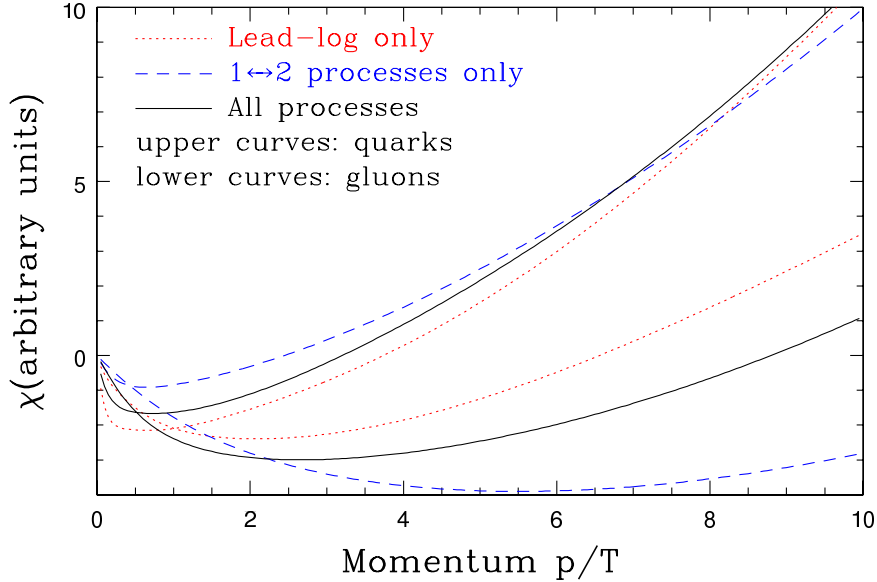}
\caption{\label{fig4} (Color online)
Functional form of $\chi(p)$ as a function of $p$,
shown for quarks and gluons in massless $\nf=3$ QCD at $\alphas=.053$
(or $\md/T=1$).  The three curves are the functional form using the
leading-log $2\leftrightarrow 2$ processes only, using the
number changing processes only, and using all processes.}
\end{figure}

For bulk
viscosity, on the other hand, the charge $q^a$ changes sign as a
function of the particle's momentum, as the $1/p$ and $p$ terms in
\Eq{eq:q_is} change relative importance.  The $1/p$ term is also
larger for gluons than for quarks, due to their larger thermal masses;
therefore, over most of the momentum range the quarks and gluons display
opposite departures from equilibrium.  In QCD, the physics of bulk viscosity is
primarily the physics of re-arranging the $|\p|$ dependence of particle
distributions.  This is what the number changing $1 \leftrightarrow 2$
processes do best; so they play a much larger role in bulk viscosity
than in shear.  Indeed, unlike the case of shear, dropping the
$2\leftrightarrow 2$ processes and retaining only the number changing
ones would still give a finite answer for $\zeta$---which in fact turns
out to be within a factor of 2 of the leading-order answer over most of
the range of $\alphas$ we have considered.  However, the detail of how
they re-arrange the momentum distributions is different than for the
elastic processes.  Therefore the detailed $p$ dependence of $\chi(p)$
is quite different if only the leading-log $2\leftrightarrow 2$
processes are considered, than if the full collision integral is used.
We illustrate this difference in Fig.~\ref{fig4}.
This limits the range of validity of the expansion in logs to the regime
where the $2\leftrightarrow 2$ processes are much faster.  But as we
just said, the $1\leftrightarrow 2$ processes are more important to bulk
viscosity than to shear, so this requires the logarithm actually to be
large.  Therefore the expansion in logs works poorly and should not be
used in treating bulk viscosity.

Another consequence of the quite nontrivial
form of $\chi(p)$ is that several basis functions must be used
to get accurate numerical values of $\zeta$.
For instance, we need at least 5 basis
functions to get $0.1\%$ accuracy, something accomplished with two
basis functions for shear viscosity.  For this reason, the results
presented in Fig.~\ref{fig1} and Fig.~\ref{fig2} are ``only'' good to
about $0.1\%$.

\begin{table}
\centerline{\begin{tabular}{|c|c|c|c|}\hline
QCD, $\nf=$ & Leading-log $A$ & NLL $\mu^*/T$ &
$\zeta_{1\leftrightarrow 2}/\alphas^2T^3$ \\ \hline
0  &  0.443 & 7.14 & .151 \\
2  &  0.638 & 7.57 & .282 \\
3  &  0.657 & 7.77 & .286 \\
4  &  0.650 & 7.93 & .279 \\
5  &  0.622 & 8.06 & .263 \\
6  &  0.577 & 8.17 & .242 \\ \hline
\end{tabular}}
\caption{\label{table1} Next-to-leading log bulk viscosity,
$\zeta=A\alphas^2T^3/\ln[\mu^*/\md]$, and $\zeta$ calculated using only
number changing collinear processes, $\zeta_{1\leftrightarrow 2}$.
All $\nf$ quarks are taken to be massless.}
\end{table}

\begin{figure}
\centerbox{0.7}{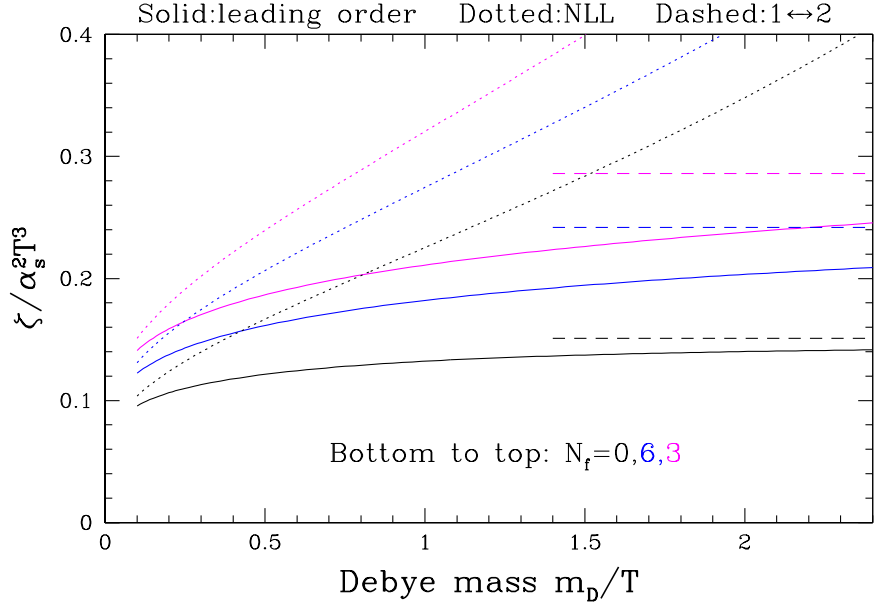}
\caption{\label{fig5} (Color online)
Bulk viscosity $\zeta$, plotted against $\md$
rather than $\alphas$ for massless QCD.
The dotted curves are the leading-log results;
the dashed line on the right is the result neglecting everything but
number changing collinear processes.}
\end{figure}

For completeness, Table \ref{table1} lists the first two coefficients in
an expansion in leading logs for massless QCD:
\begin{equation}
\zeta = \frac{A \alphas^2 T^3}{\ln[\mu^*/\md]} \, ,
\qquad
\md^2 = 2 m^2_\infty[\mbox{gluon}] = (1 + \sixth \nf) g^2 T^2 \, .
\end{equation}
The table also contains the coefficient $\zeta = C \alphas^2 T^3$ we
would obtain if we ignored {\em all} $2\leftrightarrow 2$ processes and
considered only the number changing processes.  To display the futility
of using the next-to-leading log results, we compare them with the
leading order results in Fig.~\ref{fig5}.
The failure of the next-to-leading log approximation
by a factor of at least 1.5 for
$\nf = 3$ or $6$ at $\md \ge 1.1 \, T$ corresponds to
$\alphas \ge 0.05$.


\section{Discussion}
\label{discussion}

The physics of bulk viscosity in QCD is very interesting.  The QCD
plasma leaves equilibrium under compression or rarefaction only due to
conformal symmetry breaking, and the bulk viscosity depends
quadratically on the size of conformal symmetry violation (either
through quark masses or the beta function).  To find the departure from
equilibrium one must include the forward-scattering corrections to
dispersion relations, and must account carefully for the shift in the
plasma temperature.  The departure from equilibrium due to compression
is of opposite sign for high and low momentum
excitations, and of opposite sign at intermediate momenta $p\sim \pi T$
for quarks versus gluons.  Collinear splitting processes actually
dominate the equilibration of the plasma except at very small coupling,
although in the formal weak coupling limit, equilibration should be
logarithmically
dominated by $2\leftrightarrow 2$ scattering, annihilation, and Compton
processes---with the proviso that soft gluon bremsstrahlung is also
included, since it prevents the development of a chemical potential for
particle number.  Putting this physics all together, one finds that the
bulk viscosity at leading-log order ({\it i.e.}\ for exceptionally
small coupling) is $\zeta \sim \alphas^2 T^3/\log[1/\alphas]$, with a
leading coefficient of about 1.
More practically, however, one can see from
Fig.\ \ref{fig1} that the complete result to leading order
in powers of $\alphas$ is
roughly $\zeta \simeq 0.2 \, \alphas^2 T^3$
for any reasonable perturbative value of $\alphas$
($0.02 \lesssim \alphas \lesssim 0.3$).

The practical import of bulk viscosity in QCD is very limited, however,
in the regime where a perturbative treatment has any hope of
applicability.  We find that even for $\alphas = 1/3$, the bulk
viscosity is hundreds of times smaller than the shear viscosity.  In
practice, this means that bulk viscosity can be neglected, whenever
shear viscosity plays a role.  For instance, the decay of a sound wave
depends on the combination $\zeta + 4\eta/3$; so one may drop the
$\zeta$ term to a very good approximation.  The expanding QCD plasma
in an ultra-high energy heavy ion collision is expected to be quite
anisotropic, so shear viscosity again plays a role and bulk viscosity
can be ignored.  Similarly, while the expansion of the QCD plasma in the
early universe should have been nearly isotropic, any flows in the
presence of a phase interface---the only circumstances where
nonequilibrium behavior may leave records in the early universe---are
expected to be quite anisotropic, and again shear viscosity will be more
important than bulk.

Besides the quite elegant physics involved in the bulk viscosity of QCD,
it also provides a nice example of the dangers of interpreting scalar field
theory as a toy model for gauge theory, with $\lambda$ playing the role
of $g^2$.
In massless $\lambda \phi^4$ theory, Jeon and Yaffe
showed \cite{JeonYaffe,Jeon} that the shear viscosity behaves as
$\eta \sim T^3/\lambda^2$, while bulk viscosity behaves as
$\zeta \sim \lambda T^3 \log^2[1/\lambda]$.  For shear viscosity, the
scalar theory provides a successful toy model, missing only the
logarithmic dependence.  For bulk viscosity, although some of the
physics is the same, scalar field theory is a misleading guide to
gauge theory, getting even the power of the coupling wrong.
The difference arises because
number-changing processes in scalar theory are slow compared to
processes which redistribute hard momenta
(rate $\lambda^3 T$ vs.\ $\lambda^2 T$); in QCD, they are fast
($\alphas^{3/2}T$ vs.\ $\alphas^2 T$).

One consequence of slow particle number changing rates for scalar
theory, observed by Jeon and Yaffe, was that the bulk viscosity did
not match the crude estimate
\begin {equation}
   \zeta \approx 15 \eta(\third - v_{\rm s}^2)^2
\label {eq:crudezeta}
\end {equation}
that had previously been made for scalar theory in the literature
\cite{Horsley}.%
\footnote{
   A similar estimate was made by
   Ref.\ \cite{Hosoya} but differs by a factor of 2.
   The difference is likely due
   to the incorrect identification of the shear viscosity $\eta$, by a
   factor of 2, in Eq.\ (2.39) of Ref.\ \cite{Hosoya}.
}
(This same relation was found earlier by Weinberg for a photon gas
coupled to hot matter \cite{Weinberg}.)
However, these same estimates turn out to
be {\it parametrically}\/ correct for QCD, reproducing (\ref{eq:param_result}).
In QCD, the bottleneck rate
is the same for both shear and bulk viscosity,
\begin {equation}
   v_{\rm s}^2 - \third = O\big(\beta(g^2)\big) + O(m_0^2/T^2)
\end {equation}
is a
measure of the deviation from conformal symmetry,
and this deviation is squared, just as discussed in section \ref{sec:basic}.
One could
reproduce (\ref{eq:crudezeta}) from the derivation of bulk viscosity
in this paper and of shear viscosity in Ref.\ \cite{lead-order} by
(i) keeping only the $|\Delta v_{\rm s}^2|$ term in the source
term (\ref{eq:q_is}),
and (ii) making a relaxation-time approximation of the collision
operator as a rate $\Gamma(\p)$
that is the same for bulk viscosity and shear viscosity.%
\footnote{
  Specifically,
  the ratio of the sources in the two cases then becomes
  $q_\zeta/q_\eta = |\Delta v_{\rm s}^2|$.
  The relaxation-time approximation is
  $[{\cal C} \chi](\p) = \Gamma(\p) \, \chi(\p)$.
  Using $\zeta = 2 Q_{\rm max} \bigl|_{l=0,q=q_\zeta}$ from
  (\ref{eq:zetaQmax}) and
  $\eta = \frac{2}{15} Q_{\rm max} \bigl|_{l=2,q=q_\eta}$ from
  Ref.\ \cite{lead-order}, one then obtains
  $\zeta/\eta = 15 q_\zeta^2 / q_\eta^2 = 15 |\Delta v_{\rm s}^2|^2$.
}
In Fig.\ \ref{fig:ratio}, the estimate (\ref{eq:crudezeta}) is shown
by the
dashed line for the leading-order result (\ref{eq:deltavs}) for
$\third - v_{\rm s}^2$.
It does reasonably well at estimating the order of magnitude of our result
for bulk viscosity.

It is interesting that there are certain strongly-coupled but
nearly conformal theories which find a parametrically
different dependence on $\third - v_{\rm s}^2$ than
(\ref{eq:crudezeta}).  In certain theories with gravity duals that
make them amenable to calculation,%
\footnote{
  For other bulk viscosity results in strongly interacting
  theories,
  see Ref.\ \cite{other_strong_eta}.
}
Ref.\ \cite{strong_eta} finds
$\zeta \sim \eta \, \big(\third - v_{\rm s}^2\big)$.
This result is difficult to understand from the picture of viscosity
developed in weakly-coupled field theories and provides an interesting
conceptual puzzle for understanding bulk viscosity in
strongly-coupled but nearly-conformal theories.


\begin{acknowledgments}

We would like to thank Larry Yaffe and Sangyong Jeon for useful
conversations.
We also thank Larry Yaffe and
Andrei Starinets for goading us into finally doing the
calculation.
This work was supported, in part,
by the U.S. Department of Energy under Grant No.~DE-FG02-97ER41027,
by the National Sciences and Engineering
Research Council of Canada, and by le Fonds Nature et Technologies du
Qu\'ebec.

\end {acknowledgments}


\appendix

\section{Orthogonality of \boldmath$\S$ to the energy zero mode}
\label{app:orthogonal}

In this appendix, we verify that
the source derived in this paper, given by
(\ref{eq:Sdef}) and (\ref{eq:q}), is orthogonal to the energy-changing
zero mode $\chi(p) \propto E_\p$ discussed in section \ref{sec:zeromodes}.
Specifically, we show that $({\cal S}, E_\p) = 0$ at the order of our
calculation.  That is,
\begin {equation}
   \sum_a \nu_a \int \frac{d^3 \p}{(2\pi)^3} \,
   f_0 (1 \pm f_0)
   \left( \frac{\p \cdot \v_\p}{3}
          - v_{\rm s}^2 \frac{\partial(\beta E_\p)}{\partial\beta} \right)
   E_\p = 0 .
\end {equation}
This can be checked directly using the QCD-specific formulas
of Sec.\ \ref{sec:qspecific}, but it is instructive to give a more
general argument.

Use $\partial f_0 = - f_0 (1 \pm f_0) \partial (\beta E_\p)$
and $v_\p = \grad_\p E_\p$ to rewrite the orthogonality condition as
\begin {equation}
   \sum_a \nu_a \int \frac{d^3 \p}{(2\pi)^3} \,
   \frac{T}{3} \, E_\p \, \p \cdot \grad_\p f_0
   =
   v_{\rm s}^2
   \sum_a \nu_a \int \frac{d^3 \p}{(2\pi)^3} \,
   E_\p \,\partial_\beta f_0 .
\label {eq:ortho}
\end {equation}
We then need the following two, slightly subtle equilibrium
relations, which we will discuss below:
\begin {eqnarray}
   \partial_\beta P &=& 
   \sum_a \nu_a \int \frac{d^3 \p}{(2\pi)^3} \,
   \frac{T}{3} \, E_\p \, \p \cdot \grad_\p f_0 ,
\label {eq:dPdb}
\\
   \partial_\beta \epsilon &=&
   \sum_a \nu_a \int \frac{d^3 \p}{(2\pi)^3} \,
   E_\p \,\partial_\beta f_0 .
\label {eq:dedb}
\end {eqnarray}
The orthogonality relation is then equivalent to the equilibrium
relation
\begin {equation}
   \partial_\beta P - v_{\rm s}^2 \partial_\beta \epsilon = 0 ,
\end {equation}
which is satisfied because
$v_{\rm s}^2 = dP/d\epsilon = (\partial_\beta P)/(\partial_\beta\epsilon)$.

For the rest of this appendix, we will use the short-hand notation
$\int$ to stand for $\sum_a \nu_a \int (d^3p)/(2\pi)^3$.

Deriving general relations for pressure and energy
density and their derivatives in a gas of quasi-particles
is slightly subtle because the effective energies $E_\p$ of
the quasi-particles depend on temperature and include the effects
of interactions with other quasi-particles.  The energy density is
not simply $\epsilon = \int E_\p f_0$,
for example, because this expression
suffers the usual Hartree problem of double-counting the interaction
energy.  [And, if $\epsilon$ actually were $\int E_\p f_0$,
we would not get (\ref{eq:dedb}) because
there would be an additional term where the $\partial_\beta$ hit
the $E_\p$.]  As discussed in Refs.\ \cite{PhiDerivable1,PhiDerivable2},
one simple solution to
this problem is to start with the entropy density
$S$ rather than $P$ or $\epsilon$.
Up to higher-order corrections which we shall review in a moment, the
entropy density
of a quasi-particle gas {\it is}\/ given by the naive ideal gas
formula,
\begin {equation}
   S = S_{\rm ideal} = \beta (P_{\rm ideal} + \epsilon_{\rm ideal})
     = \beta \int \left(\third \p\cdot\v_\p + E_\p\right) f_0
     = - \beta \int \third E_\p \, \p\cdot \grad_\p f_0 ,
\label {eq:entropy}
\end {equation}
where the last step follows by integrating the term involving
$v_\p = \grad_\p E_\p$ by parts.  Starting from this formula for
the entropy, we can then use the thermodynamic relation
$S = \partial_T P$ to write $\partial_\beta P = - T^2 S$ and obtain
(\ref{eq:dPdb}).

To get the formula for $\partial_\beta\epsilon$, it is convenient to
first use $v_\p f_0 = \mp T \grad_\p \ln(1 \pm f_0)$ and
integrate by parts to rewrite (\ref{eq:entropy}) as
\begin {equation}
   S = \int \left[\pm \ln(1 \pm f_0) + \beta E_\p f_0\right] .
\label {eq:entropy2}
\end {equation}
Then use the the thermodynamic relations $\epsilon = TS - P$
and $\partial_\beta P = -T^2 S$ to write
\begin {equation}
  \partial_\beta \epsilon = \partial_\beta (TS) - \partial_\beta P
  = T \partial_\beta S .
\end {equation}
Use of (\ref{eq:entropy2}) for $S$
then produces the desired formula (\ref{eq:dedb}).

It remains only to discuss the approximations that have been used in
this analysis.  In evaluating the entropy, the treatment of the system
as an ideal gas of on-shell propagating quasi-particles breaks down at order
$g^3$ and above.
(See, for instance, the analysis in Ref. \cite{PhiDerivable2}.)
But it is adequate to obtain the $O(g^2)$ and the $O(\mo^2)$ terms in the
entropy.  For massless QCD, that might sound inadequate, because the
breaking of conformal invariance is an $O(g^4)$ effect.  For example,
the effective energy of a hard quark is given by
\begin {equation}
   E_\p^2 \simeq p^2 + \third \, g^2(T) \, T^2
   = p^2 + \third \, g^2(\mu) \, T^2
         + \third \beta_0 \, g^4(\mu) \, T^2 \ln(T/\mu) + \cdots,
\end {equation}
and it is the last term which breaks conformal invariance.
However, this
$O(g^4)$ conformal-breaking log is determined by knowledge of the
$O(g^2)$ contribution; any $O(g^3)$ or additional $O(g^4)$ contributions
to thermodynamic quantities will be
conformal, up to corrections of $O(g^5)$,
and so will not contribute to the
leading-order bulk viscosity.



\end{document}